# Evaluating the Impacts of Transmission Expansion on Sub-Synchronous Resonance Risk

Farshid Salehi, Parimal Saraf, Azade Brahman, Mehriar A. Tabrizi
Power System Planning
DNV GL Energy Advisory, North America

*Abstract*— While transmission expansions are planned to have positive impact on reliability of power grids, they could increase the risk and severity of some of the detrimental incidents in power grid mainly by virtue of changing system configuration, consequently electrical distance. This paper aims to evaluate and quantify the impact of transmission expansion projects on Sub-Synchronous Resonance (SSR) risk through a two-step approach utilizing outage count index and Sub-synchronous damping index. A graph-theory based SSR screening tool is introduced to quantify the outage count associated with all grid contingencies which results in radial connection between renewable generation resources and nearby series compensated lines. Moreover, a frequency-scan based damping analysis is performed to assess the impact of transmission expansion on the system damping in sub-synchronous frequency range. The proposed approach has been utilized to evaluate the impact of recently-built transmission expansion project on SSR risk in a portion of Electric Reliability Council of Texas (ERCOT) grid.

*Index Terms*— Graph Theory, SSR Risk Assessment, Transmission Expansion, Topology Screening, Frequency Scan, EMT Simulations,

## I. INTRODUCTION

Modern power systems are becoming increasingly susceptible to sub-synchronous resonance (SSR) due to increased penetration of power electronic based renewable energy sources along with the installation of series compensated lines for increasing the power transfer capability along transmission paths [1]-[8]. Electric Reliability Council of Texas (ERCOT) grid is an example wherein series compensated transmission lines have been utilized to increase power transfer from Texas Panhandle region with significant penetration of the wind and solar resources into the load dominant regions. Without reliable risk identification and mitigation, SSR can cause considerable damage to transmission elements and generation resources, as occurred in past in the Texas grid [1]. As a result, many independent system operators have developed planning procedures to address and mitigate SSR risk during the planning stage of generation interconnection [7], [8]. As an example, ERCOT Nodal Protocol [7], requires, prior to the synchronization, SSR Risk Assessment to be performed to evaluate the impact of the facility addition (e.g. generation addition, transmission expansion) on the neighboring grid from SSR standpoint.

Briefly, a typical SSR Risk Assessment is expected to consist of the following three main phases, topology screening, frequency scan and detailed EMT simulation.

SSR risk is maximum when the generation resource radially connects to a series compensated lines in the system. Therefore, during topology screening phase, all contingencies with outage count less than or equal to a predetermined threshold that result in radial connection between the generator and any of the series compensated lines are shortlisted. Apart from manual computation approach for these outage counts which could be time consuming for a highly interconnected region with large number of generations, other methodologies have also been proposed in literature. In [9] and [10], a topology screening methodology has been presented in which contingencies resulting in the shortest path (in terms of the number of buses) between the generation Point of Interconnection (POI) and the series compensated lines is determined using the Ford-Fulkerson max-flow min-cut theorem. However, this approach provides the contingency set only for the shortest path neglecting other paths that might satisfy the underlying planning criteria. Furthermore, since the SSR risk essentially depends on the impedance of these radial paths, shortest path might not necessarily be the least impedance path. This paper presents a novel holistic topology screening tool that uses depth first search (DFS) algorithm to determine all radial paths from the POI to the series compensated line/s. The algorithm is also able to incorporate cycles into the results. The topology screening is followed by brief description of the frequency scanning techniques and EMT simulations under different contingencies and system conditions [11],[12].

Transmission expansions could have mitigating or deteriorating impact from SSR standpoint. They can increase the SSR risk by reducing the electrical proximity with series compensated lines or they can reduce the SSR risk by virtue of increasing the outage count required to put a generation facility in a radial connection with series compensated lines. In this paper, two indices are proposed to evaluate the impact of transmission expansion on SSR risk during a project planning phase. The first index is the number of outage counts for contingencies leading to radial connection between the generation units as series compensated lines. The outage count index represents the possibility of occurrence of a radial condition. The higher the number of outage count, the lesser the risk of SSR condition. Above-mentioned graph-theory based approach is proposed to obtain all the contingencies leading to radial connections as well as corresponding outage counts.

The sub-synchronous damping of the system is considered as the second criteria, this step comprises of damping analysis for entire range of sub-synchronous frequency for the contingencies with highest likelihood of SSR risk, i.e. with lowest outage count. In this paper, the cumulative resistance at the Point of Interconnection (POI) at resonant frequency is defined as SSR damping index. This index is calculated before and after a transmission expansion project. The smaller the negative SSR damping, the more the severity and likelihood of SSR condition.

The rest of the paper is organized as follows: Section II explains the SSR risk assessment methodology. Section III presents a case study demonstrating the application of SSR risk assessment methodology presented in this paper on a portion of a practical large-scale system. Section V provides the conclusions.

## II. SSR RISK ASSESSMENT

*A) Contingency Outage Count Analysis*

The first step towards SSR risk assessment is to perform a topology screening to identify radial connections from the POI of the generation resources to the series compensated lines as well as identify the contingencies resulting in those radial connections. DNV GL has developed a topology screening tool based on a modified version of a popular graph theoretic technique known as depth first search (DFS) that identifies all the radial paths (without any cycles) between the POI of the generation and either end of a series compensated line up to a depth *n* [13]. Therefore, only paths of length less than *n* are returned. The value of *n* based on the experience with smaller test systems and practical large-scale systems was selected to be 15. DFS is a graph traversal algorithm that starts at the source node and explores the edges along each path stemming from the source node till it reaches the deepest level of that path before backtracking [14]. The entire topology screening stage has been explained below to determine radial connections between the source node (POI) and the sink node (terminal bus of a series compensated line):

- DFS algorithm is used to determine all radial paths from the source node to the sink node. Multiple buses along each radial path can have branches connected to them that terminate in generation, shunts or tertiary winding of a transformer. These branches are not considered towards the N-14 contingency criteria and are therefore removed. All other branches connected to buses in each of the radial path that are not contained in that path are disconnected and form the contingency set for that particular radial path.

- All the buses that form part of each of the radial paths from the source node to sink node along with their neighboring buses are used to define a new reduced graph (subgraph of the original graph) known as *network subgraph*. This is done to reduce the degree of the graph to aid the computational speed for future steps. All paths originating from source node and terminating in sink node involve only a few buses. The neighboring buses are included in order to generate the list of contingencies or the branches that have to disconnected to obtain each radial path.

- Next important step is to identify cycles while traversing from the source node to the sink node. The importance of determining cycles for SSR risk assessment is twofold:

  1) Since SSR risk is dependent on the impedance of the radial connection, cycles in the path result in lesser impedance as compared to individual circuits.
  2) The contingency count with cycle/s in path is invariably lesser as compared to individual paths to achieve a radial connection between the source node and the sink node.

  The first step towards incorporating cycles in the paths while traversing from source node to the sink node is to find out all the cycles in the *network subgraph*. The cycles in the *network subgraph* are determined using the "cycle-basis" functionality of the NetworkX toolbox in python which is based on [15]. This algorithm determines minimal collection of cycles in a graph such that any cycle in the network can be written as a sum of cycles in the basis.

Once the information regarding all the cycles in the *network subgraph* as well as the radial paths between the source node and sink node is available, the cycles in the path can be determined. Each edge/branch in each of the extracted radial paths determined in the first step is investigated to be a part of a cycle. If an edge/branch in a particular path is contained within one/multiple cycles, that edge is replaced by the cycle/s. As mentioned, this approach is followed for each of the edge/branch that would result in redundant cycles. These redundant cycles are represented as a unique cycle. This will be explained in detail in section III.

This explains the entire topology screening tool required as a first step towards SSR risk assessment. The tool has been developed in Python using the NetworkX toolbox. The only inputs to the tool are the source node, sink node (one of the terminals of the series compensated line) and the series compensated line of interest. The output of the tool is all the radial connections between the source node and the sink node (with and without the cycles in between) as well the associated contingency set and outage rank. If any of the radial connections meets the N-14 contingency criteria, it is shortlisted for frequency scan detailed in the next section. The tool has been tested on smaller test cases as well as practical large-scale systems. In all the scenarios, the tool provides the output in less than 5 seconds.

*B) Sub-Synchronous Damping Analysis*

The impact of transmission expansion on Sub-synchronous damping can be evaluated through frequency scan. The proposed approach includes performing frequency scan for all contingencies to extract the resistance of system as function of frequency for entire range of sub-synchronous frequency. This analysis should be performed before and after transmission upgrades. If the transmission system side damping analysis are indicative of decrease in sub-synchronous resistance, further detail damping analysis is suggested for nearby generation resources. The detail damping analysis includes performing comprehensive frequency scan on both generation resources and transmission system. To make sure that the detail study covers all scenario of operations and expected sub-synchronous range, a sensitivity analysis is performed around different dispatch level and number of in service turbines/inverters of generation resources. On the other side, for the transmission system a sensitivity analysis is considered around the status of critical switch shunts.

The cumulative resistance of Point of Interconnection (POI) at resonant frequency is calculated for all above-mentioned scenario. The cumulative resistance is defined as the Sub-synchronous damping index, the smaller the negative damping the higher chance of SSR. This index is required to be calculated before and after transmission expansion to comment on the impact of project on SSR condition.

## III. CASE STUDY AND RESULTS

Texas Panhandle region with significant penetration of the wind generation resources that exports the generated power through the series compensated lines is selected to test the proposed SSR risk assessment approach. This portion of the ERCOT grid is depicted in Fig. 1. This portion of grid has recently experienced an upgrade through the addition of a second circuit shown using dashed red lines in Fig. 1

To evaluate the impact of this transmission upgrade on SSR phenomena, a 150MW wind farm connected to the station #3 is considered and both topology screening and detail SSR analysis are performed. Using the developed topology screening tool, the contingency list that results in radial connection between the station #3 and station #1 as well as station #3 and Boundary Bus#3 are extracted before and after the transmission system development.

DFS algorithm determines all radial paths from station #3 to station #1 as well as Boundary Bus#3. Following this step, all the buses in all the extracted paths as well as their immediately neighboring buses are used to form the *network subgraph*. Then, the "cycle-basis" function of the NetworkX toolbox determines the following cycles in the *network subgraph*: Station #1-Station #2-Station #3-Station #4-Station #5-Station #1 (*cycle 1*), Station #1- Station #5-Station #6-Station #7-Station #2-Station #1 (*cycle 2*), Station #1- Station #5- Station #7-Station #2-Station #1 (*cycle 3*), Station #1-Boundary Bus#2-Boundary Bus#3-Station #9-Station #8-Station #1 (*cycle 4*) and Station #9-Boundary Bus#3-Boundary Bus#4-Boundary Bus#1-Station#9 (*cycle 5*). After this step, each of the edges/branches in all the paths are explored to be a part of any of the enlisted cycles. As an illustration, one of the radial paths extracted by the DFS algorithm traversing from Station #3 to Boundary Bus#3 is Station #3-Station #4-Station #5-Station #1-Station #8-Station #9-Boundary Bus#3. Each of the edges/branches in this path is investigated to be a part of a cycle. For e.g. branch Station #5 – Station #1 is contained in three cycles, i.e., *cycle 1, cycle 2 and cycle 3*. The branch Station #5 –Station # 1 in the radial path is replaced by *cycle 1-cycle 2-cycle 3*. Thus, the radial connection between Station #3 and Boundary Bus#3 with the cycles included is *cycle 1-cycle2* ∪ *cycle3-cycle 4-cycle 5*. In a different scenario where it is assumed that the double circuit branch between Station #9 and Boundary Bus#3 does not exist, the radial connection between Station #3 and Boundary Bus#3 would be *cycle 1-cycle2* ∪ *cycle3*-Boundary Bus#2-Boundary Bus#3.

Some of the contingency lists extracted by the topology screening tool before and after transmission upgrade (meeting the ERCOT N-14 criteria) are listed in tables I and II, respectively. The developed toolbox is able to extract the radial paths and associated contingencies on the ERCOT grid in less than 5 seconds. As it is evident from Table I and II, the contingency rank increases after the upgrade for the corresponding radial conditions. Therefore, the upgrade decreases the chance of radial condition and consequently the SSR risk. However, as it was discussed earlier, the contingency rank is not the only influential index in SSR phenomena and the changes in transmission system damping should also be taken into account. In this regard, the transmission system resistance and reactance for CTG#1 & 2 while all switch shunt between POI and series compensated line are off line is depicted in Fig. 2 for both before and after upgrades.

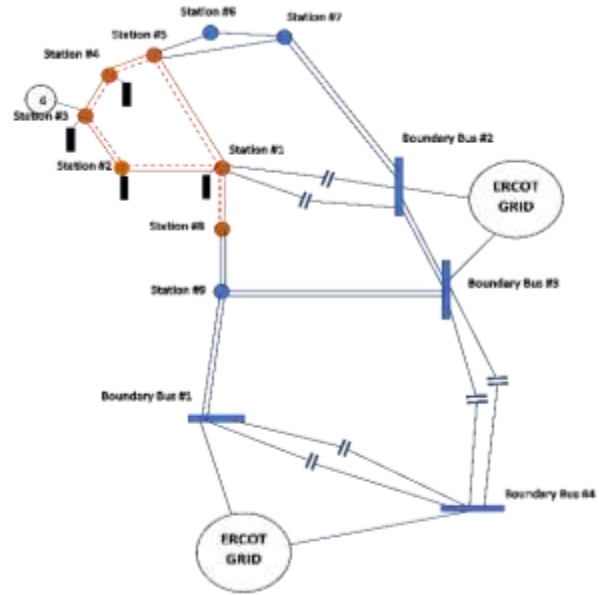

Fig. 1 Test system for SSR risk assessment

Table I: Contingency List without the upgrade

| Contingency # | From Bus Number | To Bus Number | Ckt ID | Comments |
|---|---|---|---|---|
| CTG #1 | Station #1 | Station #8 | 1 | Radial path from Station #3 to Station #1 |
| | Station #5 | Station #7 | 1 | |
| | Station #5 | Station #6 | 1 | |
| CTG #2 | Station #1 | Station #8 | 1 | Radial path from Station #3 to Station #1 |
| | Station #1 | Station #5 | 1 | |
| | Station #3 | Station #4 | 1 | |
| CTG #3 | Boundary Bus #3 | ERCOT GRID | 1 | Radial path from Station #3 to Boundary Bus#3 |
| | Boundary Bus #2 | ERCOT GRID | 1 | |
| | Station #9 | Boundary Bus#1 | 1 | |
| | Station #9 | Boundary Bus#1 | 2 | |

Table II: Contingency List with the upgrade

| Contingency # | From Bus Number | To Bus Number | Ckt ID | Comments |
|---|---|---|---|---|
| CTG #1 | Station #1 | Station #8 | 1 | Radial path from Station #3 to Station #1 |
| | Station #1 | Station #8 | 2 | |
| | Station #5 | Station #7 | 1 | |
| | Station #5 | Station #6 | 1 | |
| CTG #2 | Station #1 | Station #8 | 1 | Radial path from Station #3 to Station #1 |
| | Station #1 | Station #8 | 2 | |
| | Station #1 | Station #5 | 1 | |
| | Station #1 | Station #5 | 2 | |
| | Station #3 | Station #4 | 1 | |
| | Station #3 | Station #4 | 2 | |
| CTG#3 | Same as Table 1 | | | |

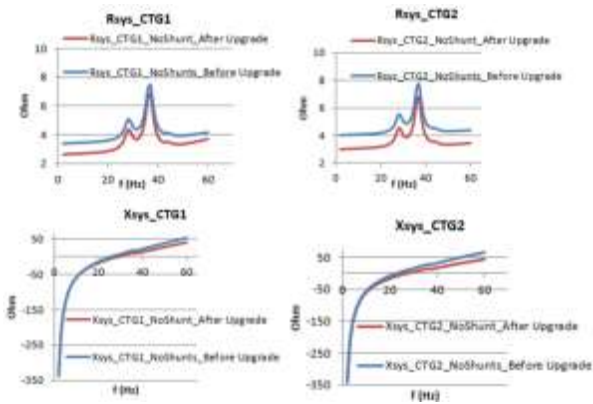

Fig. 2 System side resistance and reactance before and after upgrade

The results indicate that the upgrade lead to the decrease in system sub synchronous resistance and consequently less damping in sub-synchronous frequency range. To further corroborate the results. A detail frequency scan is performed for the proposed wind farm under contingency 1 before and after transmission system upgrade. To cover all expected operational conditions, a sensitivity analysis was performed around dispatch level of windfarm, number of turbines in service and also the status of critical switch shunt in transmission system.

The cumulative frequency scan results are depicted in Fig. 3 and 4. Also, the summary of the resonant frequency and associated damping at the POI for different scenarios of operation are provided in Table III and IV.

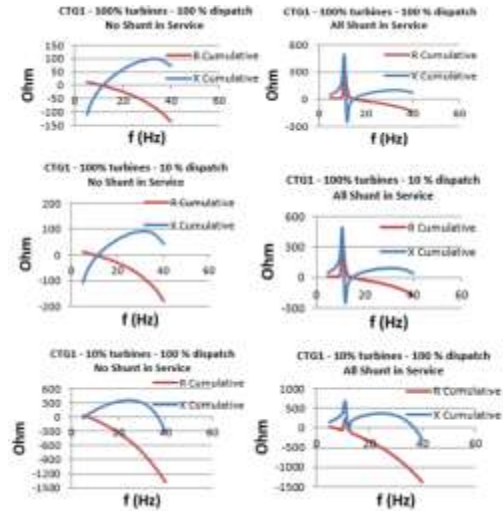

Fig. 3 Cumulative frequency scan results at POI before upgrade

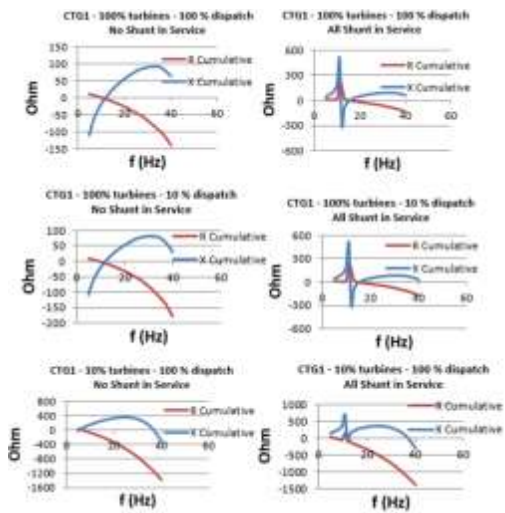

Fig. 4 Cumulative frequency scan results at POI after upgrade

Table III: Frequency scan results for different operational conditions of the wind farm and transmission systems before upgrade

| Wind Farm Operation Scenario | Transmission Side Scenario | Cross-over Frequency X (Hz) | Cumulative R (Ohm) | Observation |
|---|---|---|---|---|
| 100% turbines at 100% Dispatch | No Shunt | 11 | -3.23 | Undamped SSCI/IGE |
| | All Shunt | 15 | -8.07 | Undamped SSCI/IGE |
| 100% turbines at 10% Dispatch | No Shunt | 11 | -5.55 | Undamped SSCI/IGE |
| | All Shunt | 15 | -11.68 | Undamped SSCI/IGE |
| 10% turbines at 100% Dispatch | No Shunt | 5 | 17.27 | Damped Oscillation |
| | All Shunt | 12 | -105.23 | Undamped SSCI/IGE |

Results of frequency scan results shows:

1) Negative cumulative resistance at sub-synchronous resonate frequency represent the negative damping of the system for these modes and consequently the high risk of SSR for before and after transmission system upgrades.

Table IV: Frequency scan results for different operational conditions of the wind farm and transmission systems after upgrade

| Wind Farm Operation Scenario | Transmission Side Scenario | Cross-over Frequency X (Hz) | Cumulative R (Ohm) | Observation |
|---|---|---|---|---|
| 100% turbines at 100% Dispatch | No Shunt | 11 | -3.98 | Undamped SSCI/IGE |
| | All Shunt | 15 | -8.98 | Undamped SSCI/IGE |
| 100% turbines at 10% Dispatch | No Shunt | 11 | -6.3 | Undamped SSCI/IGE |
| | All Shunt | 15 | -12.58 | Undamped SSCI/IGE |
| 10% turbines at 100% Dispatch | No Shunt | 5 | 16.52 | Damped Oscillation |
| | All Shunt | 12 | -106.17 | Undamped SSCI/IGE |

2) Since the frequency and damping of system varies under different operational conditions the oscillation can be the result of Induction Generator Effect (IGE) or Sub-Synchronous Control Interaction (SSCI).

3) IGE and SSCI are electrical phenomena and have potential to grow rapidly, consequently leads to series damage to the transmission elements and renewable generation resources. Therefore, it is critical to correctly evaluate the risk of these phenomena for any system upgrades

4) In this case, the results of frequency scan indicate that proposed system upgrades lead to more negative damping for all scenario of operation and consequently exacerbate the SSR condition.

To validate the results, a detail EMT simulation is performed for scenario of 100% turbines in service at 100% dispatch for both all shunt in service and all shunts out of service before and after system upgrades. The active power output of the wind farm and corresponding FFTs are depicted in Fig. 5. In line with the results of frequency scan, the EMT simulation and FFT analysis show that system upgrade worsens the SSR condition by increasing the magnitude of oscillation.

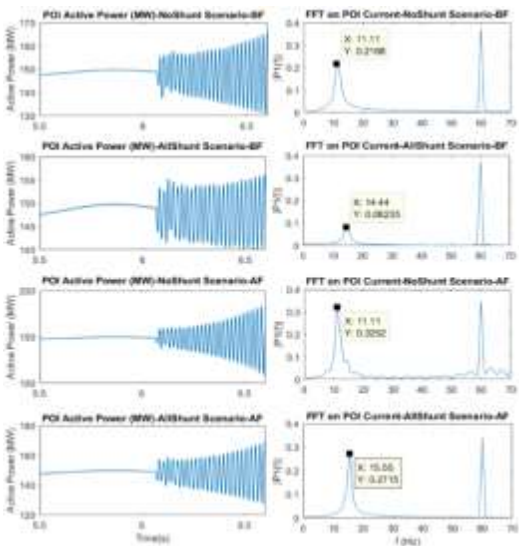

Fig. 5 EMT simulation results for no-shunt and all shunt scenarios, before and after upgrade

## IV. CONCLUSION

This paper has provided a detailed methodology for evaluating the impact of transmission expansion projects on SSR risk through a twostep approach utilizing outage count index and Sub-Synchronous damping index. A novel graph-theory based SSR screening tool has been developed and presented that determines all radial connections between the renewable generation resources and the nearby series compensated lines. Moreover, a frequency-scan based damping analysis is performed to assess the impact of transmission expansion on the system damping in sub-synchronous frequency range. The proposed methodology and developed tool were utilized on a portion of ERCOT grid and it was demonstrated that transmission expansion projects can have potential detrimental effects on SSR risk in the system.